\documentstyle[12pt]{article}
\begin{document}
\title{EXPANSION-FREE EVOLVING SPHERES MUST HAVE INHOMOGENEOUS ENERGY DENSITY DISTRIBUTIONS}
\author{ L. Herrera$^1$\thanks{e-mail: laherrera@cantv.net.ve}, G. Le
Denmat$^2$\thanks{e-mail: gerard.le$_-$denmat@upmc.fr} and  N. O.
Santos$^{2,3}$\thanks{e-mail: nilton.santos@upmc.fr} \\
\small{$^1$Escuela de F\'{\i}sica, Facultad de Ciencias,} \\
\small{Universidad Central de Venezuela, Caracas, Venezuela.}\\
\small{$^2$LERMA-PVI, Universit\'e Paris 06, Observatoire de Paris, CNRS,}\\
\small{3 rue Galil\'ee, Ivry sur Seine 94200, France.}\\
\small{$^3$Laborat\'orio Nacional de Computa\c{c}\~ao Cient\'{\i}fica,
25651-070 Petr\'opolis RJ, Brazil.}}
\maketitle

\begin{abstract}

\end{abstract}
In a recent paper  a systematic  study on shearing expansion-free spherically symmetric  distributions was presented. As a particular case of such systems, the  Skripkin model was mentioned, which corresponds to a nondissipative perfect fluid with a constant energy density. Here we show that such a model is inconsistent with junction conditions. It is shown that in general for any nondissipative fluid distribution, the expansion-free condition requires the energy density to be inhomogeneous. As an example we consider the case of dust, which allows for a complete integration.

\newpage
\section{Introduction}
In a recent paper \cite{H1} a general study on shearing expansion-free (vanishing expansion scalar $\Theta$)  spherical fluid evolution is presented, which includes pressure anisotropy and dissipation. The interest of such models stems from the fact  that  the expansion-free condition necessarily implies the appearance of a cavity.  

Indeed, it  is intuitively clear that  in the case of an overall expansion, the increase in  volume due to the increasing  area of  the external boundary surface must be compensated with  the increase of the   area of the internal boundary surface (delimiting the cavity) in order to keep $\Theta$ vanishing. The argument in the case of collapse is similar.

More rigorously, it was shown in \cite{H1} that  for  two concentric fluid shells in the process of expansion, in the neighborhood of the centre,  the $\Theta=0$ condition is violated, implying thereby that such a  condition requires that the innermost  shell of fluid should be away from the centre,  initiating therefrom  the formation of the cavity (see \cite{H1} for details).

The natural appearance of a vacuum cavity within the fluid distribution in expansion-free solutions suggests that they  might be relevant for the modelling of voids observed at cosmological scales.

In the particular case of a non-dissipative isotropic fluid, with constant energy density, the Skripkin model \cite{Skripkin} is  recovered.

The purpose of this Brief Report is twofold. We apply the results of \cite{H1} to prove that the Skripkin model is ruled out by  junction conditions. Secondly, we consider the case of dust, and provide a complete integration of this model.

\section{The expansion-free sphere}
We consider a spherically symmetric distribution  of collapsing
fluid, bounded by a spherical surface $\Sigma^{(e)}$. The fluid is
assumed to be locally anisotropic (principal stresses unequal)  but nondissipative.

Choosing comoving coordinates inside $\Sigma^{(e)}$, the general
interior metric can be written
\begin{equation}
ds^2_-=-A^2dt^2+B^2dr^2+R^2(d\theta^2+\sin^2\theta d\phi^2),
\label{1}
\end{equation}
where $A$, $B$ and $R$ are functions of $t$ and $r$ and are assumed
positive. We number the coordinates $x^0=t$, $x^1=r$, $x^2=\theta$
and $x^3=\phi$.

Outside $\Sigma^{(e)}$ we assume we have the Schwarzschild 
spacetime, described by
\begin{equation}
ds^2=-\left[1-\frac{2M}{r}\right]dv^2-2drdv+r^2(d\theta^2
+\sin^2\theta
d\phi^2) \label{1int},
\end{equation}
where $M=constant$  denotes the total mass,
and  $v$ is the retarded time.

The matching of  (\ref{1}) to the Schwarzschild spacetime (\ref{1int}) on $\Sigma^{(e)}$ requires  the continuity of the first and second differential forms, which implies  (see \cite{H1} for details)
\begin{equation}
m(t,r)\stackrel{\Sigma^{(e)}}{=}M, \qquad P_r\stackrel{\Sigma^{(e)}}{=}0,\label{junction1S}
\end{equation}

where    $\stackrel{\Sigma^{(e)}}{=}$ means that both sides of the equation
are evaluated on $\Sigma^{(e)}$,  and $m$ is the mass  function introduced by Misner and Sharp
\cite{Misner} (see also \cite{Cahill}) given by
\begin{equation}
m=\frac{R^3}{2}{R_{23}}^{23}
=\frac{R}{2}\left[\left(\frac{\dot R}{A}\right)^2-\left(\frac{R^{\prime}}{B}\right)^2+1\right].
 \label{17masa}
\end{equation}

As we mentioned in the introduction, the expansion-free models  present an internal vacuum cavity. If we call $\Sigma^{(i)} $ the boundary surface between the cavity and the fluid, then the matching of the Minkowski spacetime within the cavity to the fluid distribution, implies
\begin{equation}
m(t,r)\stackrel{\Sigma^{(i)}}{=}0, \qquad P_r\stackrel{\Sigma^{(i)}}{=}0. \label{junction1i}
\end{equation}

Now, it can be shown (see \cite{H1} for details) that as consequence of the expansion-free condition and  one of the Einstein field equations, we can write the line element as
\begin{equation}
ds^2=-\left(\frac{R^2 {\dot R}}{\tau_1}\right)^2dt^2+\frac{1}{R^4}dr^2+R^2(d\theta^2+\sin^2\theta d\phi^2),
\label{25III}
\end{equation}
where $\tau_1$ is a function of time and the dot stands for differentiation with respect to $t$ (a unit constant with dimensions $[r^4]$ is assumed to multiply $dr^2$). This is the general metric for a spherically symmetric
anisotropic perfect fluid undergoing shearing and expansion-free evolution  (observe that it  has the same form as for the isotropic fluid \cite{Mac}).

For the line element (\ref{25III}), the Einstein equation $G_{0 0}^-=8\pi T_{0 0}^-$, becomes (Eq.(76) in \cite{H1})
\begin{eqnarray}
8\pi\mu=-2R^3R^{\prime\prime}-5R^2R^{\prime 2}+\frac{1}{R^2}-3\frac{\tau^2_1}{R^6}, \label{40}\end{eqnarray}
where $\mu$ is the energy density and  the  prime stands for $r$
differentiation. The  first integral of (\ref{40}) is
\begin{equation}
R^{\prime 2}=\frac{1}{R^4}+\frac{\tau_2-2m}{R^5}+\frac{\tau^2_1}{R^8}, \label{45I}
\end{equation}
where $\tau_2(t)$ is an arbitrary function of $t$.

Using the  proper time derivative $D_T$,

\begin{equation}
D_T=\frac{1}{A}\frac{\partial}{\partial t}, \label{16}
\end{equation}
and the proper radial derivative $D_R$,
\begin{equation}
D_R=\frac{1}{R^{\prime}}\frac{\partial}{\partial r}, \label{23a}
\end{equation}
where $R$ defines the areal radius of a spherical surface inside $\Sigma^{(e)}$ (as
measured from its area), the following equations are easily obtained.  From
(\ref{17masa}),
\begin{eqnarray}
D_Tm=-4\pi P_r UR^2,
\label{22Dt}
\end{eqnarray}
where $P_r$ denotes the radial pressure and  $U$ is the velocity of the collapsing fluid,
\begin{equation}
U=D_TR<0 \;\; \mbox{(in the case of collapse)},\label{19}
\end{equation}
being the variation of the areal radius with respect to proper time. From (\ref{17masa}) too we can obtain,
\begin{eqnarray}
D_Rm=4\pi \mu R^2,
\label{27Dr}
\end{eqnarray}
implying
\begin{equation}
m=4\pi\int^{R}_{0}\mu R^2dR, \label{27intcopy}
\end{equation}
with the assumption of a regular centre to the distribution  $m(0)=0$.

From (\ref{25III}) with (\ref{19}) it follows
\begin{equation}
U=\frac{\dot R}{A}=\frac{\tau_1}{R^2}, \qquad B=\frac{1}{R^2}.
\label{uexpf}
\end{equation}
Substituting (\ref{uexpf}) into (\ref{17masa}) and using (\ref{45I}) we obtain
\begin{equation}
\tau_2=0 \label{x}.
\end{equation}

The Einstein equation $G_{11 }^-=8\pi T_{11}^-$ reads (Eq.(78) in \cite{H1})
\begin{equation}
8\pi P_r=\frac{\tau^2_1}{R^5{\dot R}}\left(3\frac{\dot R}{R}-2\frac{{\dot \tau}_1}{\tau_1}\right)
+R^3R^{\prime}\left(2\frac{{\dot R}^{\prime}}{\dot R}+5\frac{R^{\prime}}{R}\right)-\frac{1}{R^2}, \label{41}
\end{equation}
where $P_r$ denotes the radial pressure.
Using (\ref{45I}), (\ref{17masa}) and (\ref{uexpf}) in (\ref{41}) we have
\begin{equation}
8\pi P_r=-\frac{2\dot m}{R^2 \dot R}.\label{47}
\end{equation}

Observe that (\ref{47}) is fully consistent  with the junction conditions (\ref{junction1S}), (\ref{junction1i}).

Now, if we assume as Skripkin  \cite{Skripkin}, that $\mu=\mu_0=$ constant, we obtain, using (\ref{27intcopy}),
\begin{equation}
\dot m=4\pi\mu_0 R^2 \dot R.
\label{p1}
\end{equation}
Feeding back (\ref{p1}) into (\ref{47}) produces
\begin{equation}
P_r=-\mu_0=\mbox {constant},
\label{p2}
\end{equation}
which by virtue of the junction condition
\begin{equation}
P_r\stackrel{\Sigma^{(e)}}{=}0,\label{j3pi}
\end{equation}
implies
\begin{equation}
P_r=\mu_0=0.
\label{p3}
\end{equation}

Thus, the Skripkin model is ruled out by  junction conditions. Observe that the isotropy of pressure is not explicitly used. However this last condition follows from the constancy of the energy density and the expansion-free condition, as can be seen from the Bianchi identity $T^{-\alpha\beta}_{;\beta}V_{\alpha}=0$, where $V^\alpha$ denotes the four velocity of the fluid, reading (Eq.(80) in \cite{H1})
\begin{eqnarray}
{\dot\mu}+2(P_{\perp}-P_r)\frac{\dot R}{R}=0, \label{43}
\end{eqnarray}
where $P_{\perp}$ denotes the tangential pressure.
From (\ref{43}) we have that if the expansion-free fluid is isotropic, $P_r=P_{\perp}$, then the energy density $\mu$ is only $r$ dependent, and viceversa.

In the next section we consider the case of dust, $P_{\perp}=P_r=0$, with $\mu=\mu(r)$.

\section{The expansion-free dust}
The Bianchi identity $T^{-\alpha\beta}_{;\beta}\chi_{\alpha}=0$, where $\chi_{\alpha}$ is a unit four vector along the radial direction, reads (Eq.(81) in \cite{H1})
\begin{equation}
 \label{43pn}\\
P_r^{\prime}+(\mu+P_r)\frac{{\dot R}^{\prime}}{\dot R}+2(\mu+2P_r-P_{\perp})\frac{R^{\prime}}{R}=0, \label{44}
\end{equation}
and for dust it becomes
\begin{equation}
\frac{\dot R^{\prime}}{\dot R}+\frac{2 R^{\prime}}{R}=0,
\label{pnn}
\end{equation}
whose integration gives
\begin{equation}
\dot R=\frac{f(t)}{R^2}, \qquad R^{\prime}=\frac{g(r)}{R^2},
\label{pnn1}
\end{equation}
with $f(t)$ and $g(r)$ denoting arbitrary functions of their arguments.
Then from (\ref{pnn1}) we obtain
\begin{equation}
R^3=\psi(t)+\chi(r),
\label{pnn2}
\end{equation}
with
\begin{equation}
\psi(t)=3\int f(t) dt \qquad \chi(r)=3\int g(r)dr.
\label{pnn3}
\end{equation}
Without loss of generality we may choose $\tau_1(t)=f(t)$, implying, because of (\ref{25III}), (\ref{uexpf}) and (\ref{pnn1}),
\begin{equation}
A=1,\qquad U=\dot R.
\label{pnn4}
\end{equation}
Then, from the  junction condition (\ref{junction1i}), using (\ref{25III}) and (\ref{17masa})  we obtain
\begin{equation}
{\dot R}^2\stackrel{\Sigma^{(i)}}{=}g-1, \label{junction1ippn}
\end{equation}
producing
\begin{equation}
R\stackrel{\Sigma^{(i)}}{=}(g-1)^{1/2}t+R(0).\label{junction1ippn5}
\end{equation}
Evaluating  (\ref{pnn2}) at $\Sigma^{(i)}$ and considering (\ref{junction1ippn5}) we obtain
\begin{equation}
\psi(t)\stackrel{\Sigma^{(i)}}{=}\left[(g-1)^{1/2}t+R(0)\right]^3-\chi,\label{junction1ippn6}
\end{equation}
thereby providing the explicit time dependence of the models.

In order to find the $r$ dependence ($\phi$ or $\chi$) we proceed as follows.
Because of  (\ref{43}) we know that $\mu=\mu(r)$, then evaluating (\ref{40}) at $t=0$, we obtain a differential equation for $\phi(r)$ (or $\chi(r)$), which may be integrated for any  given function $\mu=\mu(r)$.

\section{Conclusions}
We have seen so far that  expansion-free condition together with junction conditions rule out the Skripkin model (constant energy density and isotropic pressure). In principle there could be constant energy density models, if we allow for the presence of dissipation, however we do not know at this point if  such models may satisfy  the whole set of junction and physical conditions.

Next, we consider the case of dust  with $\mu=\mu(r)$. These models can be completely integrated for any given function $\mu(r)$. Of course such models are members of  the Lema\^{\i}tre-Tolman-Bondi (LTB) spacetimes \cite{lemaitre}-\cite{bondi}.

Indeed, the general metric \cite{sussman} for these spacetimes,
\begin{equation}
ds^2=-dt^2+\frac{R^{\prime 2}}{1-K}dr^2+R^2(d\theta^2+\sin^2\theta d\phi^2),
\label{25IIIppnj}
\end{equation}
appears to be identical to (\ref{25III}) with (\ref{pnn1}), (\ref{pnn4})  and the identification $g^2=1-K$.

Before concluding we would like to emphasize once again that the main appeal of the expansion--free models resides in the unavoidable  appearance of a vacuum cavity  within the fluid distribution. 

This fact  suggests  that such models  might be used to describe the formation of voids observed at cosmological scales (see \cite{lid},  \cite{voids} and references therein) for  very different kinds of fluid distributions (dust, anisotropic fluids, and  dissipative fluids).

Thus, in the dust case discussed above, analytical solutions  are available,  which are relatively
simple to analyze but still may contain some of the essential features of a
realistic situation.

\section*{Acknowledgments.}
LH wishes to acknowledge financial support from the
FUNDACION EMPRESAS POLAR,  the  CDCH at Universidad Central
de Venezuela under grants PG 03-00-6497-2007 and PI 03-00-7096-2008 and Universit\'e Pierre et Marie Curie (Paris).

\end{document}